\documentclass{article}
\usepackage{latexsym}
\usepackage{authblk}
\usepackage{lingmacros,xypic,prooftree,amssymb,lscape,a4wide}
\usepackage{txfonts}
\usepackage{stmaryrd}
\usepackage{wasysym}

\usepackage{doc}
\usepackage{makeidx}

\newcommand{\easychair}{\textsf{easychair}}
\newcommand{\miktex}{MiK{\TeX}}
\newcommand{\texniccenter}{{\TeX}nicCenter}
\newcommand{\makefile}{\texttt{Makefile}}
\newcommand{\latexeditor}{LEd}

\newcommand{\defn}[2]{\enumsentence{
{\bf Definition} ({\em #1})\\\ \\#2}}

\newcommand{\thm}[2]{\enumsentence{
{\bf Theorem} ({\em #1})\\\ \\#2}}

\newcommand{\corol}[2]{\enumsentence{
{\bf Corollary} ({\em #1})\\\ \\#2}}

\newcommand{\disp}[1]{\enumsentence{#1}}

\newcommand{\tb}{\hspace*{0.25in}}
\newcommand{\tab}{\hspace*{0.5in}}

\newcommand{\yields}{\mbox{\ $\Rightarrow$\ }}

\newcommand{\bsl}{\mbox{$\backslash$}}
\newcommand{\product}{\mbox{$\bullet$}}

\newcommand{\sinfix}[1]{\mbox{$\downarrow_{#1}$}}
\newcommand{\smwrap}[1]{\mbox{$\,|{_{#1}}\,$}}
\newcommand{\mwrap}[1]{\mbox{$|$}}
\newcommand{\scircum}[1]{\mbox{$\uparrow_{#1}$}}
\newcommand{\swprod}[1]{\mbox{$\odot{_{#1}}$}}
\newcommand{\D}{\mbox{\bf D}}

\newcommand{\DbC}{\mbox{$\mathbf{Db\univexp}$}}
\newcommand{\DbCM}{\mbox{$\mathbf{Db\univexp\exstexp}$}}
\newcommand{\DbCb}{\mbox{$\mathbf{Db\univexp_b}$}}
\newcommand{\DbCbMr}{\mbox{$\mathbf{Db\univexp_b\exstexp_r}$}}
\newcommand{\bydef}{\mbox{$\stackrel{def}{=}$}} 
\newcommand{\degcon}{\mbox{$d_c$}}
\newcommand{\lexord}{\mbox{$\prec_{\mbox{\scriptsize{Lex-}}\mathbb{N}^2}$}}
\newcommand{\FLC}{\mbox{$\mathbf{FLC}$}}
\newcommand{\FLexpC}{\mbox{$\mathbf{FLC}\univexp$}}
\newcommand{\LL}{\mbox{$\mathbf{LL}$}}

\newcommand{\prf}[1]{\noindent{\bf Proof}. #1 $\square$}
\newcommand{\abrack}{\mbox{$[\,]^{-1}$}}
\newcommand{\mybrack}{\mbox{$\langle\rangle$}}
\newcommand{\sep}{\mbox{$1$}}
\newcommand{\vect}[1]{\overrightarrow{#1}}

\newcommand{\univexp}{\mbox{!}}
\newcommand{\exstexp}{\mbox{?}}

\newcommand{\ass}{\mbox{$:\,$}}

\newcommand{\subst}[1]{\mbox{$\{#1\}$}}
\newcommand{\zero}{\mbox{$0$}}
\newcommand{\commentout}[1]{}

\usepackage{natbib}

\newcommand{\syncnst}[1]{\mbox{\it #1}}
\newcommand{\CN}{\mbox{$\mathit{CN}$}}

\newcommand{\VP}{\mbox{$\mathit{VP}$}}

\author[1]{Glyn Morril}
\author[1]{Oriol Valent\'in}
\affil[1]{Department of Computer Science, Universitat Polit\`ecnica de Catalunya}

\title{Computational Coverage of TLG: Nonlinearity\thanks{Research partially supported by
SGR2014-890 (MACDA) of the Generalitat de Catalunya,
MICINN project BASMATI (TIN2011-27479-C04-03) and 
MINECO project APCOM (TIN2014-57226-P).}}

\begin{document}
\maketitle
%


%





\begin{abstract}

We study nonlinear connectives (exponentials) in the context of Type Logical Grammar
(TLG). We devise four conservative extensions of the
displacement calculus with brackets, \DbC, \DbCM, \DbCb{} and \DbCbMr{} which contain the universal and existential exponential modalities of linear logic (\LL{}). These modalities
do not exhibit the same structural properties as in \LL, which in TLG are especially adapted for linguistic purposes. The universal modality \univexp{}
for TLG allows only the commutative and contraction rules, but not weakening, whereas the existential modality \exstexp{} allows the so-called (intuitionistic) Mingle rule, which 
derives a restricted version of weakening. We provide a Curry-Howard labelling for both exponential connectives. As it turns out,
controlled contraction by \univexp{} gives a way to account for the so-called parasitic gaps, and controlled Mingle \exstexp{} iteration, in particular iterated
coordination. Finally, the four calculi are proved to be Cut-Free, and decidability is proved for a linguistically sufficient special case of $\DbCbMr$ (and hence \DbCb{}).
\end{abstract}

\setcounter{tocdepth}{2}
{\small
\tableofcontents}

%
%

\pagestyle{empty}

\section{Introduction}
Categorial logic such as displacement calculus~\D{}~\cite{mvf:tdc} is intuitionistic sublinear logic. A major innovation of linear logic
are the so-called exponentials which afford a controlled use of structural rules. Here we look at linguistically relevant exponentials 
in TLG: a universal exponential without weakening in relation to parasitic gaps,
and a restriction of the existential exponential to mingle in relation to iterated coordination:
\disp{
\begin{tabular}[t]{ll}
a.&man who$_i$ the friends of $t_i$ admire $t_i$ without praising $t_i$\\
b. & John praises, likes, and will love London.
\end{tabular} 
}
In Section~\ref{DbC} we define two logically simple calculi \DbC{} and \DbCM{}
with Curry-Howard labelling and we discuss their linguistic suitability. In section~\ref{DbCb} we define linguistically 
refined versions \DbCb{} and \DbCbMr{}, improving the previous calculi in respect of capturing the `parasicity' of parasitic
gaps, that is that, seemingly, parasitic gaps must appear in islands. In Section~\ref{Cutdec} we discuss Cut-elimination
and decidability.

\section{Db extended with contraction and mingle modalities}

\label{DbC}

The displacement calculus with brackets $\mathbf{Db}$ is defined in Figures~\ref{slcmult}, \ref{sldmult} and~\ref{slbrmod}. 
The calculus \DbC{} is obtained by adding the universal exponential rules in Figure~\ref{slunivexp}.
We denote $\DbCM$ the universal exponential displacement calculus with, in addition, the existential exponential rules of Figure~\ref{slexstexp}.

\begin{figure}[ht]
\begin{center}\footnotesize
$
\begin{array}{lc}
1. &\prooftree
\Gamma\yields B\ass\psi \tb
\Delta\langle\vect{C}\ass z\rangle\yields D\ass\omega
\justifies
\Delta\langle\vect{C/B}\ass x, \Gamma\rangle\yields D\ass\omega\subst{(x\ \psi)/z}
\using / L
\endprooftree \tb
\prooftree
\Gamma, \vect{B}\ass y\yields C\ass\chi
\justifies
\Gamma\yields C/B\ass\lambda y\chi
\using / R
\endprooftree
\\\\
2. &
\prooftree
\Gamma\yields A\ass\phi \tb
\Delta\langle\vect{C}\ass z\rangle\yields D\ass\omega
\justifies
\Delta\langle\Gamma, \vect{A\bsl C}\ass y\rangle\yields D\ass\omega\subst{(y\ \phi)/z}
\using \bsl L
\endprooftree \tb
\prooftree
\vect{A}\ass x, \Gamma\yields C\ass\chi
\justifies
\Gamma\yields A\bsl C\ass\lambda x\chi
\using \bsl R
\endprooftree
\\\\
3. & \prooftree
\Delta\langle\vect{A}\ass x, \vect{B}\ass y\rangle\yields D\ass\omega
\justifies
\Delta\langle\vect{A\product B}\ass z\rangle\yields D\ass\omega\subst{\pi_1 z/x, \pi_2 z/y}
\using \product L
\endprooftree \tb
\prooftree
\Gamma_1\yields A\ass\phi\tb\Gamma_2\yields B\ass\psi
\justifies
\Gamma_1, \Gamma_2\yields A\product B\ass(\phi, \psi)
\using \product R
\endprooftree
\\\\
4. &
\prooftree
\Delta\langle\Lambda\rangle\yields A\ass\phi
\justifies
\Delta\langle\vect{I}\ass x\rangle\yields A\ass\phi
\using IL
\endprooftree\tb
\prooftree
\justifies
\Lambda\yields I\ass\zero
\using IR
\endprooftree
\end{array}
$
\end{center}
\caption{Semantically labelled continuous multiplicative rules}
\label{slcmult}
\end{figure}
\begin{figure}[ht]
\begin{center}\footnotesize
$
\begin{array}{lc}
5. &
\prooftree
\Gamma\yields B\ass\psi \tb
\Delta\langle\vect{C}\ass z\rangle\yields D\ass\omega
\justifies
\Delta\langle\vect{C\scircum{k} B}\ass x\smwrap{k}\Gamma\rangle\yields D\ass\omega\subst{(x\ \psi)/z}
\using \scircum{k} L
\endprooftree \tb
\prooftree
\Gamma\smwrap{k}\vect{B}\ass y\yields C\ass\chi
\justifies
\Gamma\yields C\scircum{k} B\ass\lambda y\chi
\using \scircum{k} R
\endprooftree
\\\\
6. & 
\prooftree
\Gamma\yields A\ass\phi \tb
\Delta\langle\vect{C}\ass z\rangle\yields D\ass\omega
\justifies
\Delta\langle\Gamma\smwrap{k}\vect{A\sinfix{k} C}\ass y\rangle\yields D\ass\omega\subst{(y\ \phi)/z}
\using \sinfix{k} L
\endprooftree \tb
\prooftree
\vect{A}\ass x\smwrap{k}\Gamma\yields C\ass\chi
\justifies
\Gamma\yields A\sinfix{k} C\ass\lambda x\chi
\using \sinfix{k} R
\endprooftree
\\\\
7. &
\prooftree
\Delta\langle\vect{A}\ass x\smwrap{k}\vect{B}\ass y\rangle\yields D\ass\omega
\justifies
\Delta\langle\vect{A\swprod{k} B}\ass z\rangle\yields D\ass\omega\subst{\pi_1 z/x, \pi_2 z/y}
\using \swprod{k} L
\endprooftree \tb
\prooftree
\Gamma_1\yields A\ass\phi\tb\Gamma_2\yields B\ass\psi
\justifies
\Gamma_1\smwrap{k}\Gamma_2\yields A\swprod{k} B\ass(\Phi,\Psi)
\using \swprod{k} R
\endprooftree
\\\\
8. &
\prooftree
\Delta\langle\sep\rangle\yields A\ass\phi
\justifies
\Delta\langle\vect{J}\ass x\rangle\yields A\ass\phi
\using JL
\endprooftree\tb
\prooftree
\justifies
\sep\yields J\ass\zero
\using JR
\endprooftree
\end{array}
$
\end{center}
\caption{Semantically labelled discontinuous multiplicative rules}
\label{sldmult}
\end{figure}

\begin{figure}[ht]
\begin{center}\footnotesize
$
\begin{array}{lc}
15. & \prooftree
\Delta\langle \vect{A}\ass x\rangle\yields B\ass\psi
\justifies
\Delta\langle[\vect{\abrack A}\ass x]\rangle\yields B\ass\psi
\using \abrack L
\endprooftree \tb
\prooftree
[\Gamma]\yields A\ass\phi
\justifies
\Gamma\yields \abrack A\ass\phi
\using \abrack R
\endprooftree
\\\\
16. & \prooftree
\Delta\langle[\vect{A}\ass x]\rangle\yields B\ass\psi
\justifies
\Delta\langle\vect{\mybrack A}\ass x\rangle\yields B\ass\psi
\using \mybrack L
\endprooftree \tb
\prooftree
\Gamma\yields A\ass\phi
\justifies
[\Gamma]\yields \mybrack A\ass\phi
\using \mybrack R
\endprooftree
\end{array}$
 \end{center}
\caption{Semantically labelled bracket modality rules}
\label{slbrmod}
\end{figure}

\begin{figure}[ht]
\begin{center}\footnotesize
$
\begin{array}{lc}
17. &  \prooftree
 \Gamma\langle A\ass x\rangle\yields B\ass\psi
 \justifies
 \Gamma\langle\univexp A\ass x\rangle\yields B\ass\psi
 \using \univexp L
 \endprooftree 
 \tb
\prooftree
 \univexp A_1\ass x_1, \ldots, \univexp A_n\ass x_n \yields A\ass\phi
 \justifies
 \univexp A_1\ass x_1, \ldots, \univexp A_n\ass x_n\yields \univexp A\ass\phi
 \using \univexp R
 \endprooftree
 \\\\
& \prooftree
 \Delta\langle\univexp A\ass x, \Gamma\rangle\yields B\ass\psi
 \justifies
 \Delta\langle\Gamma, \univexp A\ass x\rangle\yields B\ass\psi
 \using \univexp P
 \endprooftree 
 \tb
 \prooftree
 \Delta\langle\Gamma, \univexp A\ass x\rangle\yields B\ass\psi
 \justifies
 \Delta\langle\univexp A\ass x, \Gamma\rangle\yields B\ass\psi
 \using \univexp P
 \endprooftree 
 \\\\
 & \prooftree
 \Delta\langle\univexp A_0\ass x_0, \ldots, \univexp A_n\ass x_n, \univexp A_0\ass y_0, \ldots,
 \univexp A_n\ass y_0\rangle\yields B\ass\psi
 \justifies
 \Delta\langle\univexp A_0\ass x_0, \ldots, \univexp A_n\ass x_n\rangle\yields B\ass\psi\subst{x_0/y_0, \ldots, x_n/y_n}
 \using \univexp C
 \endprooftree 
\end{array}
$
 \end{center}
\caption{Semantically labelled universal exponential rules}
\label{slunivexp}
\end{figure}


\begin{figure}[ht]
\begin{center}\footnotesize
$
\begin{array}{lc}
 18. &
\prooftree
\univexp\Gamma(A\ass x)\yields \exstexp B\ass \psi(x)
\justifies
\univexp\Gamma(\exstexp A\ass z)\yields \exstexp B\ass\bigoplus_{x\in z}\psi(x)
\using \exstexp L
\endprooftree
\tb
\prooftree
\Gamma\yields A\ass\phi
\justifies
\Gamma\yields \exstexp A\ass[\phi]
\using \exstexp R
\endprooftree
\\\\
&\prooftree
\Gamma\yields \exstexp A\ass\phi\tb\Delta\yields \exstexp A\ass\psi
\justifies
\Gamma, \Delta\yields \exstexp A\ass\phi\oplus\psi
\using \exstexp M
\endprooftree
\end{array}
$
 \end{center}
\caption{Semantically labelled existential exponential rules}
\label{slexstexp}
\end{figure}

The very elementary characterisation of (object) relativisation is obtained by assigning a relative pronoun type
$(\CN\bsl\CN)/(S/N)$. This captures the long distance character of relativisation but only allows peripheral extraction. 
Using the universal exponential we can improve the type assignment to $(\CN\bsl\CN)/$ $(S/\univexp N)$ which,
in view of the permutability of the exponential subtype also allows medial extraction.

Various `islands' can inhibit or block relativisation:
weak islands such as subjects (Chomsky 1973\cite{chomsky:73}) and adverbial phrases,
from which extraction is mildly unacceptable,
and strong islands such as coordinate structures (Ross 1967\cite{ross})
and relative clauses themselves,
from which extraction is entirely unacceptable:
\disp{\begin{tabular}[t]{ll}
a. & ?man who$_i$ the friend of $t_i$ laughed\\
b. & ?paper which$_i$ John laughed before reading $t_i$
\end{tabular}}
\disp{\begin{tabular}[t]{ll}
a. & *man who$_i$ John laughed and Mary likes $t_i$\\
b. & *man who$_i$ John likes the woman that loves $t_i$
\end{tabular}}

Furthermore,
relativisation can also comprise `parasitic extraction' in which a relative
pronoun binds more than one extraction site (Taraldsen 1979\cite{taraldsen:79};
Engdahl 1983\cite{engdahl:83};
Sag 1983\cite{sag:gaps}).
There must be a  `host' gap which is not in an island,
and according to the received wisdom, and according with the terminology `parasitic', this may license a `parasitic' gap in (any number of immediate
weak) islands:
\disp{\begin{tabular}[t]{ll}
a. & the man who$_i$ the friends of $t_i$ admire $t_i$\\
b. & the paper which$_i$ John filed $t_i$ without reading $t_i$\\
c. & the paper which$_i$ the editor of $t_i$ filed $t_i$ without reading $t_i$
\end{tabular}}
In addition, we observe that
these parasitic gaps may in turn function as host gaps licensing
further parasitic gaps in (weak) subislands,
and so on recursively:
\disp{\begin{tabular}[t]{ll}
a. & man who$_i$ the fact that the friends of $t_i$ admire $t_i$ 
surprises $t_i$\\
b. &  man who$_i$ the fact that the friends of $t_i$ admire $t_i$ without praising $t_i$
offends $t_i$ without surprising $t_i$
\end{tabular}}

The bracket modalities of Figure~\ref{slbrmod} have application to syntactical domains
such as prosodic phrases and extraction islands.
For example,
$\syncnst{walks}\ass\mybrack{}N\bsl S$ for the subject condition,
and
$\syncnst{before}\ass$ $\abrack{}(\VP\bsl\VP)/\VP$ for the adverbial island constraint. 
The relative pronoun type $(\CN\bsl\CN)/(S/\univexp N)$ respects these island constraints because the brackets
induced block association and permutation of the exponential hypothetical subtype into the bracketed domains.

The presence of the contraction rule potentially allows for parasitic extraction, but in fact the islands in which the parasitic gaps
are supposed to occur are closed off for the reasons just given. Furthermore the calculus as it stands overgenerates pseudo-parasitic
multiple extraction in which `parasitic' gaps do not occur in islands:
\disp{
\begin{tabular}[t]{ll}
a.& * the slave who$_i$ John sold $t_i$ to $t_i$\\
b.&  * the slave who$_i$ John sold $t_i$ $t_i$ 
\end{tabular}
}
Thus the logic of contraction as it stands precisely both undergenerates and overgenerates parasitic extraction. We fix this in the next section.

Using the existential exponential, 
\exstexp,
we can assign a coordinator type
$\syncnst{and}\ass(\exstexp N\bsl N)/N$ allowing iterated coordination
as in $\syncnst{John, Bill, Mary\ and\ Suzy}\ass N$,
or $\syncnst{and}\ass(\exstexp (S/N)\bsl (S/N))/${}$(S/N)$ for
$\syncnst{John\ likes,\ Mary}$ $\syncnst{dislikes, and\ Bill\ hates, London}$ (iterated right node raising),
and so on.

\section{Db extended with restricted modalised 
contraction and mingle}

\label{DbCb}

The calculus \DbCb{} is obtained by adding to $\bf Db$ the restricted universal exponential rules in Figure~\ref{slunivexpb}.
Note how now the application of contraction induces a bracketed domain.
We denote $\DbCbMr$ the restricted universal exponential displacement calculus with, in addition, the existential exponential restricted to only succedent occurrences,
and with only the rules of Figure~\ref{slexstexpb}.

\begin{figure}[ht]
\begin{center}\footnotesize
$
\begin{array}{lc}
17. &  \prooftree
 \Gamma\langle A\ass x\rangle\yields B\ass\psi
 \justifies
 \Gamma\langle\univexp A\ass x\rangle\yields B\ass\psi
 \using \univexp L
 \endprooftree 
 \tb
\prooftree
 \univexp A_1\ass x_1, \ldots, \univexp A_n\ass x_n \yields A\ass\phi
 \justifies
 \univexp A_1\ass x_1, \ldots, \univexp A_n\ass x_n\yields \univexp A\ass\phi
 \using \univexp R
 \endprooftree
 \\\\
& \prooftree
 \Delta\langle\univexp A\ass x, \Gamma\rangle\yields B\ass\psi
 \justifies
 \Delta\langle\Gamma, \univexp A\ass x\rangle\yields B\ass\psi
 \using \univexp P
 \endprooftree 
 \tb
 \prooftree
 \Delta\langle\Gamma, \univexp A\ass x\rangle\yields B\ass\psi
 \justifies
 \Delta\langle\univexp A\ass x, \Gamma\rangle\yields B\ass\psi
 \using \univexp P
 \endprooftree 
 \\\\
 & \prooftree
 \Delta\langle\univexp A_0\ass x_0, \ldots, \univexp A_n\ass x_n, [\univexp A_0\ass y_0, \ldots,
 \univexp A_n\ass y_0, \Gamma]\rangle\yields B\ass\psi
 \justifies
 \Delta\langle\univexp A_0\ass x_0, \ldots, \univexp A_n\ass x_n, \Gamma\rangle\yields B\ass\psi\subst{x_0/y_0, \ldots, x_n/y_n}
 \using \univexp C_b
 \endprooftree 
\end{array}
$
 \end{center}
\caption{Semantically labelled restricted universal exponential rules}
\label{slunivexpb}
\end{figure}

\begin{figure}[ht]
\begin{center}\footnotesize
$
\begin{array}{lc}
 18. &
\prooftree
\Gamma\yields A\ass\phi
\justifies
\Gamma\yields \exstexp A\ass[\phi]
\using \exstexp R
\endprooftree
\\\\
&\prooftree
\Gamma\yields A\ass\phi\tb\Delta\yields \exstexp A\ass\psi
\justifies
\Gamma, \Delta\yields \exstexp A\ass[\phi|\psi]
\using \exstexp M_r
\endprooftree
\end{array}
$
 \end{center}
\caption{Semantically labelled restricted existential exponential rules}
\label{slexstexpb}
\end{figure}

In the following subsections we report analyses computer-generated by
a categorial parser/theorem-prover CatLog2.

\subsection{Parasitic relativisation}

As we have remarked
subjects are weak islands;
accordingly in our CatLog fragment there is no derivation of simple relativization
from a subject such as:
\disp{
${\bf man}{+}[[{\bf that}{+}[{\bf the}{+}{\bf friends}{+}{\bf of}]{+}{\bf walk}]]: {\it CN}{\it s(m)}$}
(Note the strong island double brackets of the relative
clause ensuring that it is an island from which parasitic extraction is not possible.)
However, a weak island `parasitic' gap 
can be licensed by a host gap 
\cite{taraldsen:79}:
\disp{
${\bf man}{+}[[{\bf that}{+}{\bf the}{+}{\bf friends}{+}{\bf of}{+}{\bf admire}]]: {\it CN}{\it s(m)}$}
Lexical lookup yields:
\disp{
${\square}{\it CN}{\it s(m)}: {\it man}, [[{\blacksquare}{\forall}n({[]^{-1}}{[]^{-1}}({\it CN}{\it n}\backslash {\it CN}{\it n})/{\blacksquare}(({\langle\rangle}Nt(n){\sqcap}!{\blacksquare}Nt(n))\backslash Sf)): \lambda A\lambda B\lambda C[({\it B}\ {\it C})\wedge ({\it A}\ {\it C})],\\
{}{\blacksquare}{\forall}n(Nt(n)/{\it CN}{\it n}): \iota , {\square}({\it CN}{\it p}/{\it PP}{\it of}): {\it friends}, {\square}(({\forall}n({\it CN}{\it n}\backslash {\it CN}{\it n})/{\blacksquare}{\exists}bNb){\&}({\it PP}{\it of}/{\exists}aNa)): \mbox{\^{}}(\mbox{\v{}}{\it of}, \lambda D{\it D}),\\
 {\square}(({\langle\rangle}({\exists}aNa{-}{\exists}gNt(s(g)))\backslash Sf)/{\exists}aNa): \mbox{\^{}}\lambda E\lambda F({\it Pres}\ ((\mbox{\v{}}{\it admire}\ {\it E})\ {\it F}))]]\ \Rightarrow\ {\it CN}{\it s(m)}$}
There is the following derivation,
where the use of contraction, involving brackets and,
in focused proofs,
stoups,
corresponds to generating the parasitic gap:

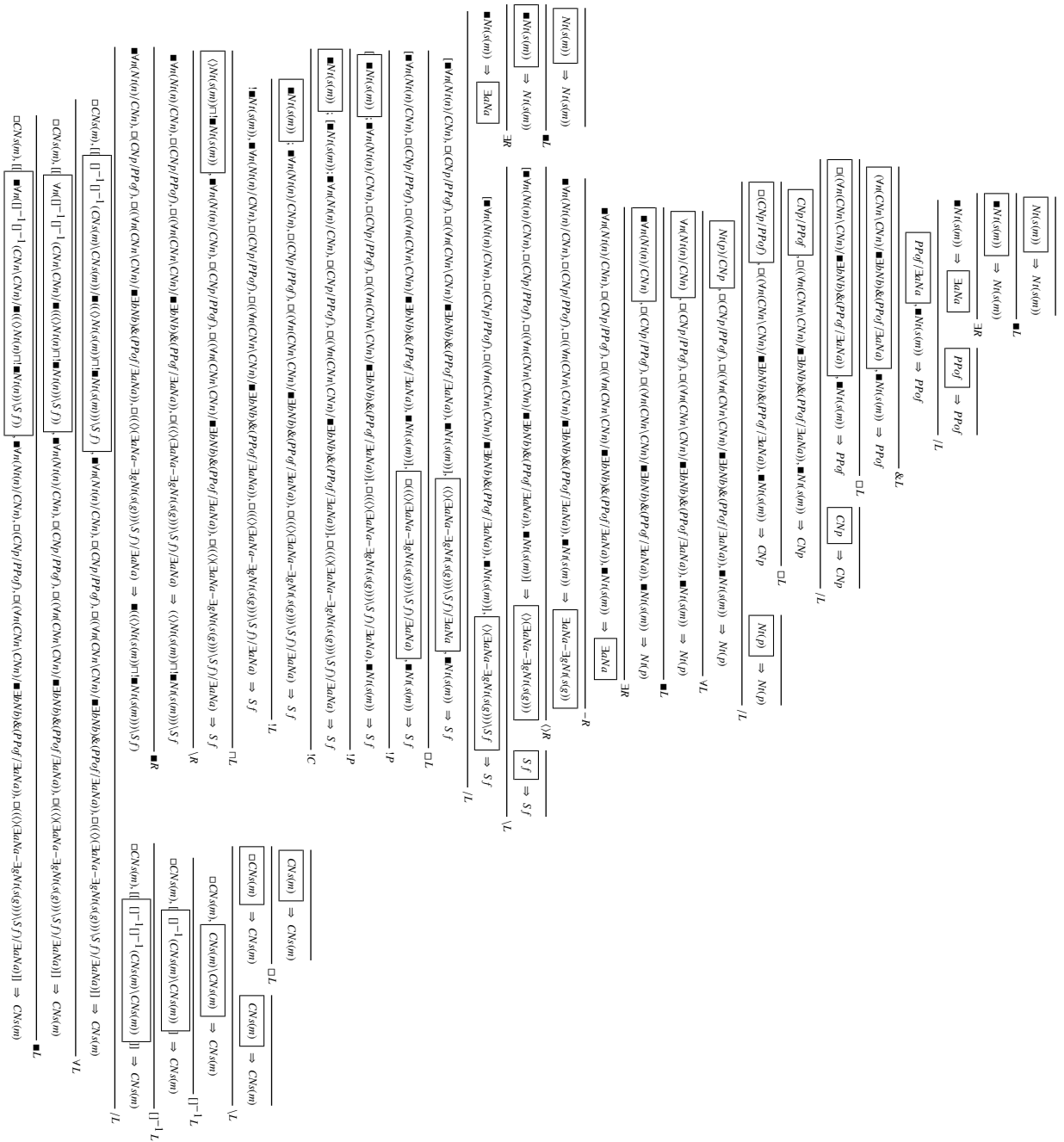
\vspace{0.15in}

\rotatebox{-90}{\tiny
\prooftree
\prooftree
\prooftree
\prooftree
\prooftree
\prooftree
\prooftree
\prooftree
\prooftree
\prooftree
\prooftree
\prooftree
\prooftree
\prooftree
\prooftree
\justifies
\mbox{\fbox{$Nt(s(m))$}}\ \Rightarrow\ Nt(s(m))
\endprooftree
\justifies
\mbox{\fbox{${\blacksquare}Nt(s(m))$}}\ \Rightarrow\ Nt(s(m))
\using {\blacksquare}L
\endprooftree
\justifies
{\blacksquare}Nt(s(m))\ \Rightarrow\ \fbox{${\exists}aNa$}
\using {\exists}R
\endprooftree
\prooftree
\prooftree
\prooftree
\prooftree
\prooftree
\prooftree
\prooftree
\prooftree
\prooftree
\prooftree
\prooftree
\prooftree
\prooftree
\prooftree
\prooftree
\justifies
\mbox{\fbox{$Nt(s(m))$}}\ \Rightarrow\ Nt(s(m))
\endprooftree
\justifies
\mbox{\fbox{${\blacksquare}Nt(s(m))$}}\ \Rightarrow\ Nt(s(m))
\using {\blacksquare}L
\endprooftree
\justifies
{\blacksquare}Nt(s(m))\ \Rightarrow\ \fbox{${\exists}aNa$}
\using {\exists}R
\endprooftree
\prooftree
\justifies
\mbox{\fbox{${\it PP}{\it of}$}}\ \Rightarrow\ {\it PP}{\it of}
\endprooftree
\justifies
\mbox{\fbox{${\it PP}{\it of}/{\exists}aNa$}}, {\blacksquare}Nt(s(m))\ \Rightarrow\ {\it PP}{\it of}
\using {/}L
\endprooftree
\justifies
\mbox{\fbox{$({\forall}n({\it CN}{\it n}\backslash {\it CN}{\it n})/{\blacksquare}{\exists}bNb){\&}({\it PP}{\it of}/{\exists}aNa)$}}, {\blacksquare}Nt(s(m))\ \Rightarrow\ {\it PP}{\it of}
\using {\&}L
\endprooftree
\justifies
\mbox{\fbox{${\square}(({\forall}n({\it CN}{\it n}\backslash {\it CN}{\it n})/{\blacksquare}{\exists}bNb){\&}({\it PP}{\it of}/{\exists}aNa))$}}, {\blacksquare}Nt(s(m))\ \Rightarrow\ {\it PP}{\it of}
\using {\Box}L
\endprooftree
\prooftree
\justifies
\mbox{\fbox{${\it CN}{\it p}$}}\ \Rightarrow\ {\it CN}{\it p}
\endprooftree
\justifies
\mbox{\fbox{${\it CN}{\it p}/{\it PP}{\it of}$}}, {\square}(({\forall}n({\it CN}{\it n}\backslash {\it CN}{\it n})/{\blacksquare}{\exists}bNb){\&}({\it PP}{\it of}/{\exists}aNa)), {\blacksquare}Nt(s(m))\ \Rightarrow\ {\it CN}{\it p}
\using {/}L
\endprooftree
\justifies
\mbox{\fbox{${\square}({\it CN}{\it p}/{\it PP}{\it of})$}}, {\square}(({\forall}n({\it CN}{\it n}\backslash {\it CN}{\it n})/{\blacksquare}{\exists}bNb){\&}({\it PP}{\it of}/{\exists}aNa)), {\blacksquare}Nt(s(m))\ \Rightarrow\ {\it CN}{\it p}
\using {\Box}L
\endprooftree
\prooftree
\justifies
\mbox{\fbox{$Nt(p)$}}\ \Rightarrow\ Nt(p)
\endprooftree
\justifies
\mbox{\fbox{$Nt(p)/{\it CN}{\it p}$}}, {\square}({\it CN}{\it p}/{\it PP}{\it of}), {\square}(({\forall}n({\it CN}{\it n}\backslash {\it CN}{\it n})/{\blacksquare}{\exists}bNb){\&}({\it PP}{\it of}/{\exists}aNa)), {\blacksquare}Nt(s(m))\ \Rightarrow\ Nt(p)
\using {/}L
\endprooftree
\justifies
\mbox{\fbox{${\forall}n(Nt(n)/{\it CN}{\it n})$}}, {\square}({\it CN}{\it p}/{\it PP}{\it of}), {\square}(({\forall}n({\it CN}{\it n}\backslash {\it CN}{\it n})/{\blacksquare}{\exists}bNb){\&}({\it PP}{\it of}/{\exists}aNa)), {\blacksquare}Nt(s(m))\ \Rightarrow\ Nt(p)
\using {\forall}L
\endprooftree
\justifies
\mbox{\fbox{${\blacksquare}{\forall}n(Nt(n)/{\it CN}{\it n})$}}, {\square}({\it CN}{\it p}/{\it PP}{\it of}), {\square}(({\forall}n({\it CN}{\it n}\backslash {\it CN}{\it n})/{\blacksquare}{\exists}bNb){\&}({\it PP}{\it of}/{\exists}aNa)), {\blacksquare}Nt(s(m))\ \Rightarrow\ Nt(p)
\using {\blacksquare}L
\endprooftree
\justifies
{\blacksquare}{\forall}n(Nt(n)/{\it CN}{\it n}), {\square}({\it CN}{\it p}/{\it PP}{\it of}), {\square}(({\forall}n({\it CN}{\it n}\backslash {\it CN}{\it n})/{\blacksquare}{\exists}bNb){\&}({\it PP}{\it of}/{\exists}aNa)), {\blacksquare}Nt(s(m))\ \Rightarrow\ \fbox{${\exists}aNa$}
\using {\exists}R
\endprooftree
\justifies
{\blacksquare}{\forall}n(Nt(n)/{\it CN}{\it n}), {\square}({\it CN}{\it p}/{\it PP}{\it of}), {\square}(({\forall}n({\it CN}{\it n}\backslash {\it CN}{\it n})/{\blacksquare}{\exists}bNb){\&}({\it PP}{\it of}/{\exists}aNa)), {\blacksquare}Nt(s(m))\ \Rightarrow\ \fbox{${\exists}aNa{-}{\exists}gNt(s(g))$}
\using {-}R
\endprooftree
\justifies
[{\blacksquare}{\forall}n(Nt(n)/{\it CN}{\it n}), {\square}({\it CN}{\it p}/{\it PP}{\it of}), {\square}(({\forall}n({\it CN}{\it n}\backslash {\it CN}{\it n})/{\blacksquare}{\exists}bNb){\&}({\it PP}{\it of}/{\exists}aNa)), {\blacksquare}Nt(s(m))]\ \Rightarrow\ \fbox{${\langle\rangle}({\exists}aNa{-}{\exists}gNt(s(g)))$}
\using {\langle\rangle}R
\endprooftree
\prooftree
\justifies
\mbox{\fbox{$Sf$}}\ \Rightarrow\ Sf
\endprooftree
\justifies
[{\blacksquare}{\forall}n(Nt(n)/{\it CN}{\it n}), {\square}({\it CN}{\it p}/{\it PP}{\it of}), {\square}(({\forall}n({\it CN}{\it n}\backslash {\it CN}{\it n})/{\blacksquare}{\exists}bNb){\&}({\it PP}{\it of}/{\exists}aNa)), {\blacksquare}Nt(s(m))], \mbox{\fbox{${\langle\rangle}({\exists}aNa{-}{\exists}gNt(s(g)))\backslash Sf$}}\ \Rightarrow\ Sf
\using {\backslash}L
\endprooftree
\justifies
[{\blacksquare}{\forall}n(Nt(n)/{\it CN}{\it n}), {\square}({\it CN}{\it p}/{\it PP}{\it of}), {\square}(({\forall}n({\it CN}{\it n}\backslash {\it CN}{\it n})/{\blacksquare}{\exists}bNb){\&}({\it PP}{\it of}/{\exists}aNa)), {\blacksquare}Nt(s(m))], \mbox{\fbox{$({\langle\rangle}({\exists}aNa{-}{\exists}gNt(s(g)))\backslash Sf)/{\exists}aNa$}}, {\blacksquare}Nt(s(m))\ \Rightarrow\ Sf
\using {/}L
\endprooftree
\justifies
[{\blacksquare}{\forall}n(Nt(n)/{\it CN}{\it n}), {\square}({\it CN}{\it p}/{\it PP}{\it of}), {\square}(({\forall}n({\it CN}{\it n}\backslash {\it CN}{\it n})/{\blacksquare}{\exists}bNb){\&}({\it PP}{\it of}/{\exists}aNa)), {\blacksquare}Nt(s(m))], \mbox{\fbox{${\square}(({\langle\rangle}({\exists}aNa{-}{\exists}gNt(s(g)))\backslash Sf)/{\exists}aNa)$}}, {\blacksquare}Nt(s(m))\ \Rightarrow\ Sf
\using {\Box}L
\endprooftree
\justifies
[\mbox{\fbox{${\blacksquare}Nt(s(m))$}};{\blacksquare}{\forall}n(Nt(n)/{\it CN}{\it n}), {\square}({\it CN}{\it p}/{\it PP}{\it of}), {\square}(({\forall}n({\it CN}{\it n}\backslash {\it CN}{\it n})/{\blacksquare}{\exists}bNb){\&}({\it PP}{\it of}/{\exists}aNa))], {\square}(({\langle\rangle}({\exists}aNa{-}{\exists}gNt(s(g)))\backslash Sf)/{\exists}aNa), {\blacksquare}Nt(s(m))\ \Rightarrow\ Sf
\using {!}P
\endprooftree
\justifies
\mbox{\fbox{${\blacksquare}Nt(s(m))$}};\ [{\blacksquare}Nt(s(m));{\blacksquare}{\forall}n(Nt(n)/{\it CN}{\it n}), {\square}({\it CN}{\it p}/{\it PP}{\it of}), {\square}(({\forall}n({\it CN}{\it n}\backslash {\it CN}{\it n})/{\blacksquare}{\exists}bNb){\&}({\it PP}{\it of}/{\exists}aNa))], {\square}(({\langle\rangle}({\exists}aNa{-}{\exists}gNt(s(g)))\backslash Sf)/{\exists}aNa)\ \Rightarrow\ Sf
\using {!}P
\endprooftree
\justifies
\mbox{\fbox{${\blacksquare}Nt(s(m))$}};\ {\blacksquare}{\forall}n(Nt(n)/{\it CN}{\it n}), {\square}({\it CN}{\it p}/{\it PP}{\it of}), {\square}(({\forall}n({\it CN}{\it n}\backslash {\it CN}{\it n})/{\blacksquare}{\exists}bNb){\&}({\it PP}{\it of}/{\exists}aNa)), {\square}(({\langle\rangle}({\exists}aNa{-}{\exists}gNt(s(g)))\backslash Sf)/{\exists}aNa)\ \Rightarrow\ Sf
\using {!}C
\endprooftree
\justifies
!{\blacksquare}Nt(s(m)), {\blacksquare}{\forall}n(Nt(n)/{\it CN}{\it n}), {\square}({\it CN}{\it p}/{\it PP}{\it of}), {\square}(({\forall}n({\it CN}{\it n}\backslash {\it CN}{\it n})/{\blacksquare}{\exists}bNb){\&}({\it PP}{\it of}/{\exists}aNa)), {\square}(({\langle\rangle}({\exists}aNa{-}{\exists}gNt(s(g)))\backslash Sf)/{\exists}aNa)\ \Rightarrow\ Sf
\using {!}L
\endprooftree
\justifies
\mbox{\fbox{${\langle\rangle}Nt(s(m)){\sqcap}!{\blacksquare}Nt(s(m))$}}, {\blacksquare}{\forall}n(Nt(n)/{\it CN}{\it n}), {\square}({\it CN}{\it p}/{\it PP}{\it of}), {\square}(({\forall}n({\it CN}{\it n}\backslash {\it CN}{\it n})/{\blacksquare}{\exists}bNb){\&}({\it PP}{\it of}/{\exists}aNa)), {\square}(({\langle\rangle}({\exists}aNa{-}{\exists}gNt(s(g)))\backslash Sf)/{\exists}aNa)\ \Rightarrow\ Sf
\using {\sqcap}L
\endprooftree
\justifies
{\blacksquare}{\forall}n(Nt(n)/{\it CN}{\it n}), {\square}({\it CN}{\it p}/{\it PP}{\it of}), {\square}(({\forall}n({\it CN}{\it n}\backslash {\it CN}{\it n})/{\blacksquare}{\exists}bNb){\&}({\it PP}{\it of}/{\exists}aNa)), {\square}(({\langle\rangle}({\exists}aNa{-}{\exists}gNt(s(g)))\backslash Sf)/{\exists}aNa)\ \Rightarrow\ ({\langle\rangle}Nt(s(m)){\sqcap}!{\blacksquare}Nt(s(m)))\backslash Sf
\using {\backslash}R
\endprooftree
\justifies
{\blacksquare}{\forall}n(Nt(n)/{\it CN}{\it n}), {\square}({\it CN}{\it p}/{\it PP}{\it of}), {\square}(({\forall}n({\it CN}{\it n}\backslash {\it CN}{\it n})/{\blacksquare}{\exists}bNb){\&}({\it PP}{\it of}/{\exists}aNa)), {\square}(({\langle\rangle}({\exists}aNa{-}{\exists}gNt(s(g)))\backslash Sf)/{\exists}aNa)\ \Rightarrow\ {\blacksquare}(({\langle\rangle}Nt(s(m)){\sqcap}!{\blacksquare}Nt(s(m)))\backslash Sf)
\using {\blacksquare}R
\endprooftree
\prooftree
\prooftree
\prooftree
\prooftree
\prooftree
\justifies
\mbox{\fbox{${\it CN}{\it s(m)}$}}\ \Rightarrow\ {\it CN}{\it s(m)}
\endprooftree
\justifies
\mbox{\fbox{${\square}{\it CN}{\it s(m)}$}}\ \Rightarrow\ {\it CN}{\it s(m)}
\using {\Box}L
\endprooftree
\prooftree
\justifies
\mbox{\fbox{${\it CN}{\it s(m)}$}}\ \Rightarrow\ {\it CN}{\it s(m)}
\endprooftree
\justifies
{\square}{\it CN}{\it s(m)}, \mbox{\fbox{${\it CN}{\it s(m)}\backslash {\it CN}{\it s(m)}$}}\ \Rightarrow\ {\it CN}{\it s(m)}
\using {\backslash}L
\endprooftree
\justifies
{\square}{\it CN}{\it s(m)}, [\mbox{\fbox{${[]^{-1}}({\it CN}{\it s(m)}\backslash {\it CN}{\it s(m)})$}}]\ \Rightarrow\ {\it CN}{\it s(m)}
\using {[]^{-1}}L
\endprooftree
\justifies
{\square}{\it CN}{\it s(m)}, [[\mbox{\fbox{${[]^{-1}}{[]^{-1}}({\it CN}{\it s(m)}\backslash {\it CN}{\it s(m)})$}}]]\ \Rightarrow\ {\it CN}{\it s(m)}
\using {[]^{-1}}L
\endprooftree
\justifies
{\square}{\it CN}{\it s(m)}, [[\mbox{\fbox{${[]^{-1}}{[]^{-1}}({\it CN}{\it s(m)}\backslash {\it CN}{\it s(m)})/{\blacksquare}(({\langle\rangle}Nt(s(m)){\sqcap}!{\blacksquare}Nt(s(m)))\backslash Sf)$}}, {\blacksquare}{\forall}n(Nt(n)/{\it CN}{\it n}), {\square}({\it CN}{\it p}/{\it PP}{\it of}), {\square}(({\forall}n({\it CN}{\it n}\backslash {\it CN}{\it n})/{\blacksquare}{\exists}bNb){\&}({\it PP}{\it of}/{\exists}aNa)), {\square}(({\langle\rangle}({\exists}aNa{-}{\exists}gNt(s(g)))\backslash Sf)/{\exists}aNa)]]\ \Rightarrow\ {\it CN}{\it s(m)}
\using {/}L
\endprooftree
\justifies
{\square}{\it CN}{\it s(m)}, [[\mbox{\fbox{${\forall}n({[]^{-1}}{[]^{-1}}({\it CN}{\it n}\backslash {\it CN}{\it n})/{\blacksquare}(({\langle\rangle}Nt(n){\sqcap}!{\blacksquare}Nt(n))\backslash Sf))$}}, {\blacksquare}{\forall}n(Nt(n)/{\it CN}{\it n}), {\square}({\it CN}{\it p}/{\it PP}{\it of}), {\square}(({\forall}n({\it CN}{\it n}\backslash {\it CN}{\it n})/{\blacksquare}{\exists}bNb){\&}({\it PP}{\it of}/{\exists}aNa)), {\square}(({\langle\rangle}({\exists}aNa{-}{\exists}gNt(s(g)))\backslash Sf)/{\exists}aNa)]]\ \Rightarrow\ {\it CN}{\it s(m)}
\using {\forall}L
\endprooftree
\justifies
{\square}{\it CN}{\it s(m)}, [[\mbox{\fbox{${\blacksquare}{\forall}n({[]^{-1}}{[]^{-1}}({\it CN}{\it n}\backslash {\it CN}{\it n})/{\blacksquare}(({\langle\rangle}Nt(n){\sqcap}!{\blacksquare}Nt(n))\backslash Sf))$}}, {\blacksquare}{\forall}n(Nt(n)/{\it CN}{\it n}), {\square}({\it CN}{\it p}/{\it PP}{\it of}), {\square}(({\forall}n({\it CN}{\it n}\backslash {\it CN}{\it n})/{\blacksquare}{\exists}bNb){\&}({\it PP}{\it of}/{\exists}aNa)), {\square}(({\langle\rangle}({\exists}aNa{-}{\exists}gNt(s(g)))\backslash Sf)/{\exists}aNa)]]\ \Rightarrow\ {\it CN}{\it s(m)}
\using {\blacksquare}L
\endprooftree}

\vspace{0.15in}

\noindent
This delivers the following semantics in which the gap variable is multiply bound:
\disp{
$\lambda C[(\mbox{\v{}}{\it man}\ {\it C})\wedge ({\it Pres}\ ((\mbox{\v{}}{\it admire}\ {\it C})\ (\iota \ (\mbox{\v{}}{\it friends}\ {\it C}))))]$}

\subsection{Iterated coordination}

To express the lexical semantics of coordination, including iterated coordination
and coordination in various arities, we use two combinators: a non-empty list
map apply $\alpha^+$ and a non-empty list map $\Phi^n$ combinator $\Phi^{n+}$.
The former is a follows:
\disp{
$
\begin{array}[t]{rcl}
(\alpha^+\ [x]\ y) & = & [(x\ y)]\\
(\alpha^+\ [x, y| z]\ w) & = & [(x\ w)|(\alpha^+\ [y|z]\ w)]
\end{array}
$}
The latter is thus:
\disp{
$
\begin{array}[t]{rcl}
(((\Phi^{n+}\ 0\ {\it and})\ x)\ [y]) & = & [y\wedge x]\\
(((\Phi^{n+}\ 0\ {\it or})\ x)\ [y]) & = & [y\vee x]\\
(((\Phi^{n+}\ 0\ {\it and})\ x)\ [y, z|w]) & = & [y\wedge(((\Phi^{n+}\ 0\ {\it and})\ x)\ [z|w])]\\
(((\Phi^{n+}\ 0\ {\it or})\ x)\ [y, z|w]) & = & [y\vee(((\Phi^{n+}\ 0\ {\it or})\ x)\ [z|w])]\\
((((\Phi^{n+}\ (s\ n)\ c)\ x)\ y)\ z) & = & (((\Phi^{n+}\ n\ c)\ (x\ z))\ (\alpha^+\ y\ z))
\end{array}
$}

Transitive verb phrase iterated coordination:
\disp{
(crd(28)) $[{\bf john}]{+}[[{\bf praises}{+}{\bf likes}{+}{\bf and}{+}{\bf will}{+}{\bf love}]]{+}{\bf london}: Sf$}
Lexical insertion yields:
\disp{
$[{\blacksquare}Nt(s(m)): {\it j}], [[{\square}(({\langle\rangle}{\exists}gNt(s(g))\backslash Sf)/{\exists}aNa): \mbox{\^{}}\lambda A\lambda B({\it Pres}\ ((\mbox{\v{}}{\it praise}\ {\it A})\ {\it B})), {\square}(({\langle\rangle}{\exists}gNt(s(g))\backslash Sf)/{\exists}aNa): \mbox{\^{}}\lambda C\lambda D({\it Pres}\ ((\mbox{\v{}}{\it like}\ {\it C})\ {\it D})), {\blacksquare}{\forall}f{\forall}a((?{\blacksquare}(({\langle\rangle}Na\backslash Sf)/{\exists}bNb)\backslash {[]^{-1}}{[]^{-1}}(({\langle\rangle}Na\backslash Sf)/{\exists}bNb))/{\blacksquare}(({\langle\rangle}Na\backslash Sf)/{\exists}bNb)): ({\Phi^n}^+\ ({\it s}\ ({\it s}\ {\it 0}))\ {\it and}), {\blacksquare}{\forall}a(({\langle\rangle}Na\backslash Sf)/({\langle\rangle}Na\backslash Sb)): \lambda E\lambda F({\it Fut}\ ({\it E}\ {\it F})), {\square}(({\langle\rangle}{\exists}aNa\backslash Sb)/{\exists}aNa): \mbox{\^{}}\lambda G\lambda H((\mbox{\v{}}{\it love}\ {\it G})\ {\it H})]],\\
 {\blacksquare}Nt(s(n)): {\it l}\ \Rightarrow\ Sf$}
The coordination combinator semantics is such that:
\disp{
$\begin{array}[t]{l}
(((((\Phi^{n+}\ (s\ (s\ 0))\ {\it and})\ x)\ [y, z])\ w)\ u) =\\
((((\Phi^{n+}\ (s\ 0)\ {\it and})\ (x\ w))\ (\alpha^+\ [y, z]\ w))\ u) =\\
((((\Phi^{n+}\ (s\ 0)\ {\it and})\ (x\ w))\ [(y\ w), (z\ w)])\ u) =\\
(((\phi^{n+}\ 0\ {\it and})\ ((x\ w)\ u))\ (\alpha^+\ [(y\ w), (z\ w)]\ u)) =\\
(((\phi^{n+}\ 0\ {\it and})\ ((x\ w)\ u))\ [((y\ w)\ u), ((z\ w)\ u)]) =\\
{}[((y\ w)\ u)\wedge[((z\ w)\ u)\wedge((x\ w)\ u)]]
\end{array}$}
There is the derivation:

\vspace{0.15in}

{\tiny

\prooftree
\prooftree
\prooftree
\prooftree
\prooftree
\prooftree
\prooftree
\prooftree
\prooftree
\prooftree
\prooftree
\prooftree
\prooftree
\prooftree
\justifies
N8344\ \Rightarrow\ N8344
\endprooftree
\justifies
N8344\ \Rightarrow\ \fbox{${\exists}aNa$}
\using {\exists}R
\endprooftree
\prooftree
\prooftree
\prooftree
\prooftree
\justifies
Nt(s(m))\ \Rightarrow\ Nt(s(m))
\endprooftree
\justifies
Nt(s(m))\ \Rightarrow\ \fbox{${\exists}aNa$}
\using {\exists}R
\endprooftree
\justifies
[Nt(s(m))]\ \Rightarrow\ \fbox{${\langle\rangle}{\exists}aNa$}
\using {\langle\rangle}R
\endprooftree
\prooftree
\justifies
\mbox{\fbox{$Sb$}}\ \Rightarrow\ Sb
\endprooftree
\justifies
[Nt(s(m))], \mbox{\fbox{${\langle\rangle}{\exists}aNa\backslash Sb$}}\ \Rightarrow\ Sb
\using {\backslash}L
\endprooftree
\justifies
[Nt(s(m))], \mbox{\fbox{$({\langle\rangle}{\exists}aNa\backslash Sb)/{\exists}aNa$}}, N8344\ \Rightarrow\ Sb
\using {/}L
\endprooftree
\justifies
[Nt(s(m))], \mbox{\fbox{${\square}(({\langle\rangle}{\exists}aNa\backslash Sb)/{\exists}aNa)$}}, N8344\ \Rightarrow\ Sb
\using {\Box}L
\endprooftree
\justifies
{\langle\rangle}Nt(s(m)), {\square}(({\langle\rangle}{\exists}aNa\backslash Sb)/{\exists}aNa), N8344\ \Rightarrow\ Sb
\using {\langle\rangle}L
\endprooftree
\justifies
{\square}(({\langle\rangle}{\exists}aNa\backslash Sb)/{\exists}aNa), N8344\ \Rightarrow\ {\langle\rangle}Nt(s(m))\backslash Sb
\using {\backslash}R
\endprooftree
\prooftree
\prooftree
\prooftree
\justifies
Nt(s(m))\ \Rightarrow\ Nt(s(m))
\endprooftree
\justifies
[Nt(s(m))]\ \Rightarrow\ \fbox{${\langle\rangle}Nt(s(m))$}
\using {\langle\rangle}R
\endprooftree
\prooftree
\justifies
\mbox{\fbox{$Sf$}}\ \Rightarrow\ Sf
\endprooftree
\justifies
[Nt(s(m))], \mbox{\fbox{${\langle\rangle}Nt(s(m))\backslash Sf$}}\ \Rightarrow\ Sf
\using {\backslash}L
\endprooftree
\justifies
[Nt(s(m))], \mbox{\fbox{$({\langle\rangle}Nt(s(m))\backslash Sf)/({\langle\rangle}Nt(s(m))\backslash Sb)$}}, {\square}(({\langle\rangle}{\exists}aNa\backslash Sb)/{\exists}aNa), N8344\ \Rightarrow\ Sf
\using {/}L
\endprooftree
\justifies
[Nt(s(m))], \mbox{\fbox{${\forall}a(({\langle\rangle}Na\backslash Sf)/({\langle\rangle}Na\backslash Sb))$}}, {\square}(({\langle\rangle}{\exists}aNa\backslash Sb)/{\exists}aNa), N8344\ \Rightarrow\ Sf
\using {\forall}L
\endprooftree
\justifies
[Nt(s(m))], \mbox{\fbox{${\blacksquare}{\forall}a(({\langle\rangle}Na\backslash Sf)/({\langle\rangle}Na\backslash Sb))$}}, {\square}(({\langle\rangle}{\exists}aNa\backslash Sb)/{\exists}aNa), N8344\ \Rightarrow\ Sf
\using {\blacksquare}L
\endprooftree
\justifies
[Nt(s(m))], {\blacksquare}{\forall}a(({\langle\rangle}Na\backslash Sf)/({\langle\rangle}Na\backslash Sb)), {\square}(({\langle\rangle}{\exists}aNa\backslash Sb)/{\exists}aNa), {\exists}bNb\ \Rightarrow\ Sf
\using {\exists}L
\endprooftree
\justifies
{\langle\rangle}Nt(s(m)), {\blacksquare}{\forall}a(({\langle\rangle}Na\backslash Sf)/({\langle\rangle}Na\backslash Sb)), {\square}(({\langle\rangle}{\exists}aNa\backslash Sb)/{\exists}aNa), {\exists}bNb\ \Rightarrow\ Sf
\using {\langle\rangle}L
\endprooftree
\justifies
{\blacksquare}{\forall}a(({\langle\rangle}Na\backslash Sf)/({\langle\rangle}Na\backslash Sb)), {\square}(({\langle\rangle}{\exists}aNa\backslash Sb)/{\exists}aNa), {\exists}bNb\ \Rightarrow\ {\langle\rangle}Nt(s(m))\backslash Sf
\using {\backslash}R
\endprooftree
\justifies
{\blacksquare}{\forall}a(({\langle\rangle}Na\backslash Sf)/({\langle\rangle}Na\backslash Sb)), {\square}(({\langle\rangle}{\exists}aNa\backslash Sb)/{\exists}aNa)\ \Rightarrow\ ({\langle\rangle}Nt(s(m))\backslash Sf)/{\exists}bNb
\using {/}R
\endprooftree
\justifies
\begin{array}{c}
{\blacksquare}{\forall}a(({\langle\rangle}Na\backslash Sf)/({\langle\rangle}Na\backslash Sb)), {\square}(({\langle\rangle}{\exists}aNa\backslash Sb)/{\exists}aNa)\ \Rightarrow\ {\blacksquare}(({\langle\rangle}Nt(s(m))\backslash Sf)/{\exists}bNb)\\
\mbox{\footnotesize\textcircled{1}}
\end{array}
\using {\blacksquare}R
\endprooftree}

\newpage

\vspace*{-1.4in}\ \\

{\tiny

\prooftree
\prooftree
\prooftree
\prooftree
\prooftree
\prooftree
\prooftree
\prooftree
\prooftree
\prooftree
\justifies
N8345\ \Rightarrow\ N8345
\endprooftree
\justifies
N8345\ \Rightarrow\ \fbox{${\exists}aNa$}
\using {\exists}R
\endprooftree
\prooftree
\prooftree
\prooftree
\prooftree
\justifies
Nt(s(m))\ \Rightarrow\ Nt(s(m))
\endprooftree
\justifies
Nt(s(m))\ \Rightarrow\ \fbox{${\exists}gNt(s(g))$}
\using {\exists}R
\endprooftree
\justifies
[Nt(s(m))]\ \Rightarrow\ \fbox{${\langle\rangle}{\exists}gNt(s(g))$}
\using {\langle\rangle}R
\endprooftree
\prooftree
\justifies
\mbox{\fbox{$Sf$}}\ \Rightarrow\ Sf
\endprooftree
\justifies
[Nt(s(m))], \mbox{\fbox{${\langle\rangle}{\exists}gNt(s(g))\backslash Sf$}}\ \Rightarrow\ Sf
\using {\backslash}L
\endprooftree
\justifies
[Nt(s(m))], \mbox{\fbox{$({\langle\rangle}{\exists}gNt(s(g))\backslash Sf)/{\exists}aNa$}}, N8345\ \Rightarrow\ Sf
\using {/}L
\endprooftree
\justifies
[Nt(s(m))], \mbox{\fbox{${\square}(({\langle\rangle}{\exists}gNt(s(g))\backslash Sf)/{\exists}aNa)$}}, N8345\ \Rightarrow\ Sf
\using {\Box}L
\endprooftree
\justifies
[Nt(s(m))], {\square}(({\langle\rangle}{\exists}gNt(s(g))\backslash Sf)/{\exists}aNa), {\exists}bNb\ \Rightarrow\ Sf
\using {\exists}L
\endprooftree
\justifies
{\langle\rangle}Nt(s(m)), {\square}(({\langle\rangle}{\exists}gNt(s(g))\backslash Sf)/{\exists}aNa), {\exists}bNb\ \Rightarrow\ Sf
\using {\langle\rangle}L
\endprooftree
\justifies
{\square}(({\langle\rangle}{\exists}gNt(s(g))\backslash Sf)/{\exists}aNa), {\exists}bNb\ \Rightarrow\ {\langle\rangle}Nt(s(m))\backslash Sf
\using {\backslash}R
\endprooftree
\justifies
{\square}(({\langle\rangle}{\exists}gNt(s(g))\backslash Sf)/{\exists}aNa)\ \Rightarrow\ ({\langle\rangle}Nt(s(m))\backslash Sf)/{\exists}bNb
\using {/}R
\endprooftree
\justifies
{\square}(({\langle\rangle}{\exists}gNt(s(g))\backslash Sf)/{\exists}aNa)\ \Rightarrow\ {\blacksquare}(({\langle\rangle}Nt(s(m))\backslash Sf)/{\exists}bNb)
\using {\blacksquare}R
\endprooftree
\prooftree
\prooftree
\prooftree
\prooftree
\prooftree
\prooftree
\prooftree
\prooftree
\prooftree
\prooftree
\justifies
N8346\ \Rightarrow\ N8346
\endprooftree
\justifies
N8346\ \Rightarrow\ \fbox{${\exists}aNa$}
\using {\exists}R
\endprooftree
\prooftree
\prooftree
\prooftree
\prooftree
\justifies
Nt(s(m))\ \Rightarrow\ Nt(s(m))
\endprooftree
\justifies
Nt(s(m))\ \Rightarrow\ \fbox{${\exists}gNt(s(g))$}
\using {\exists}R
\endprooftree
\justifies
[Nt(s(m))]\ \Rightarrow\ \fbox{${\langle\rangle}{\exists}gNt(s(g))$}
\using {\langle\rangle}R
\endprooftree
\prooftree
\justifies
\mbox{\fbox{$Sf$}}\ \Rightarrow\ Sf
\endprooftree
\justifies
[Nt(s(m))], \mbox{\fbox{${\langle\rangle}{\exists}gNt(s(g))\backslash Sf$}}\ \Rightarrow\ Sf
\using {\backslash}L
\endprooftree
\justifies
[Nt(s(m))], \mbox{\fbox{$({\langle\rangle}{\exists}gNt(s(g))\backslash Sf)/{\exists}aNa$}}, N8346\ \Rightarrow\ Sf
\using {/}L
\endprooftree
\justifies
[Nt(s(m))], \mbox{\fbox{${\square}(({\langle\rangle}{\exists}gNt(s(g))\backslash Sf)/{\exists}aNa)$}}, N8346\ \Rightarrow\ Sf
\using {\Box}L
\endprooftree
\justifies
[Nt(s(m))], {\square}(({\langle\rangle}{\exists}gNt(s(g))\backslash Sf)/{\exists}aNa), {\exists}bNb\ \Rightarrow\ Sf
\using {\exists}L
\endprooftree
\justifies
{\langle\rangle}Nt(s(m)), {\square}(({\langle\rangle}{\exists}gNt(s(g))\backslash Sf)/{\exists}aNa), {\exists}bNb\ \Rightarrow\ Sf
\using {\langle\rangle}L
\endprooftree
\justifies
{\square}(({\langle\rangle}{\exists}gNt(s(g))\backslash Sf)/{\exists}aNa), {\exists}bNb\ \Rightarrow\ {\langle\rangle}Nt(s(m))\backslash Sf
\using {\backslash}R
\endprooftree
\justifies
{\square}(({\langle\rangle}{\exists}gNt(s(g))\backslash Sf)/{\exists}aNa)\ \Rightarrow\ ({\langle\rangle}Nt(s(m))\backslash Sf)/{\exists}bNb
\using {/}R
\endprooftree
\justifies
{\square}(({\langle\rangle}{\exists}gNt(s(g))\backslash Sf)/{\exists}aNa)\ \Rightarrow\ {\blacksquare}(({\langle\rangle}Nt(s(m))\backslash Sf)/{\exists}bNb)
\using {\blacksquare}R
\endprooftree
\justifies
{\square}(({\langle\rangle}{\exists}gNt(s(g))\backslash Sf)/{\exists}aNa)\ \Rightarrow\ \fbox{$?{\blacksquare}(({\langle\rangle}Nt(s(m))\backslash Sf)/{\exists}bNb)$}
\using {?}R
\endprooftree
\justifies
\begin{array}{c}
{\square}(({\langle\rangle}{\exists}gNt(s(g))\backslash Sf)/{\exists}aNa), {\square}(({\langle\rangle}{\exists}gNt(s(g))\backslash Sf)/{\exists}aNa)\ \Rightarrow\ \fbox{$?{\blacksquare}(({\langle\rangle}Nt(s(m))\backslash Sf)/{\exists}bNb)$}\\
\mbox{\footnotesize\textcircled{2}}
\end{array}
\using {?}E
\endprooftree}

\vspace*{-0.6in}\ \\

{\tiny
\begin{center}
\rotatebox{-90}{
\prooftree
\prooftree
\prooftree
\prooftree
\mbox{\footnotesize\textcircled{1}}\tab
\prooftree
\mbox{\footnotesize\textcircled{2}}\tab
\prooftree
\prooftree
\prooftree
\prooftree
\prooftree
\prooftree
\justifies
\mbox{\fbox{$Nt(s(n))$}}\ \Rightarrow\ Nt(s(n))
\endprooftree
\justifies
\mbox{\fbox{${\blacksquare}Nt(s(n))$}}\ \Rightarrow\ Nt(s(n))
\using {\blacksquare}L
\endprooftree
\justifies
{\blacksquare}Nt(s(n))\ \Rightarrow\ \fbox{${\exists}bNb$}
\using {\exists}R
\endprooftree
\prooftree
\prooftree
\prooftree
\prooftree
\justifies
\mbox{\fbox{$Nt(s(m))$}}\ \Rightarrow\ Nt(s(m))
\endprooftree
\justifies
\mbox{\fbox{${\blacksquare}Nt(s(m))$}}\ \Rightarrow\ Nt(s(m))
\using {\blacksquare}L
\endprooftree
\justifies
[{\blacksquare}Nt(s(m))]\ \Rightarrow\ \fbox{${\langle\rangle}Nt(s(m))$}
\using {\langle\rangle}R
\endprooftree
\prooftree
\justifies
\mbox{\fbox{$Sf$}}\ \Rightarrow\ Sf
\endprooftree
\justifies
[{\blacksquare}Nt(s(m))], \mbox{\fbox{${\langle\rangle}Nt(s(m))\backslash Sf$}}\ \Rightarrow\ Sf
\using {\backslash}L
\endprooftree
\justifies
[{\blacksquare}Nt(s(m))], \mbox{\fbox{$({\langle\rangle}Nt(s(m))\backslash Sf)/{\exists}bNb$}}, {\blacksquare}Nt(s(n))\ \Rightarrow\ Sf
\using {/}L
\endprooftree
\justifies
[{\blacksquare}Nt(s(m))], [\mbox{\fbox{${[]^{-1}}(({\langle\rangle}Nt(s(m))\backslash Sf)/{\exists}bNb)$}}], {\blacksquare}Nt(s(n))\ \Rightarrow\ Sf
\using {[]^{-1}}L
\endprooftree
\justifies
[{\blacksquare}Nt(s(m))], [[\mbox{\fbox{${[]^{-1}}{[]^{-1}}(({\langle\rangle}Nt(s(m))\backslash Sf)/{\exists}bNb)$}}]], {\blacksquare}Nt(s(n))\ \Rightarrow\ Sf
\using {[]^{-1}}L
\endprooftree
\justifies
[{\blacksquare}Nt(s(m))], [[{\square}(({\langle\rangle}{\exists}gNt(s(g))\backslash Sf)/{\exists}aNa), {\square}(({\langle\rangle}{\exists}gNt(s(g))\backslash Sf)/{\exists}aNa), \mbox{\fbox{$?{\blacksquare}(({\langle\rangle}Nt(s(m))\backslash Sf)/{\exists}bNb)\backslash {[]^{-1}}{[]^{-1}}(({\langle\rangle}Nt(s(m))\backslash Sf)/{\exists}bNb)$}}]], {\blacksquare}Nt(s(n))\ \Rightarrow\ Sf
\using {\backslash}L
\endprooftree
\justifies
[{\blacksquare}Nt(s(m))], [[{\square}(({\langle\rangle}{\exists}gNt(s(g))\backslash Sf)/{\exists}aNa), {\square}(({\langle\rangle}{\exists}gNt(s(g))\backslash Sf)/{\exists}aNa), \mbox{\fbox{$(?{\blacksquare}(({\langle\rangle}Nt(s(m))\backslash Sf)/{\exists}bNb)\backslash {[]^{-1}}{[]^{-1}}(({\langle\rangle}Nt(s(m))\backslash Sf)/{\exists}bNb))/{\blacksquare}(({\langle\rangle}Nt(s(m))\backslash Sf)/{\exists}bNb)$}}, {\blacksquare}{\forall}a(({\langle\rangle}Na\backslash Sf)/({\langle\rangle}Na\backslash Sb)), {\square}(({\langle\rangle}{\exists}aNa\backslash Sb)/{\exists}aNa)]], {\blacksquare}Nt(s(n))\ \Rightarrow\ Sf
\using {/}L
\endprooftree
\justifies
[{\blacksquare}Nt(s(m))], [[{\square}(({\langle\rangle}{\exists}gNt(s(g))\backslash Sf)/{\exists}aNa), {\square}(({\langle\rangle}{\exists}gNt(s(g))\backslash Sf)/{\exists}aNa), \mbox{\fbox{${\forall}a((?{\blacksquare}(({\langle\rangle}Na\backslash Sf)/{\exists}bNb)\backslash {[]^{-1}}{[]^{-1}}(({\langle\rangle}Na\backslash Sf)/{\exists}bNb))/{\blacksquare}(({\langle\rangle}Na\backslash Sf)/{\exists}bNb))$}}, {\blacksquare}{\forall}a(({\langle\rangle}Na\backslash Sf)/({\langle\rangle}Na\backslash Sb)), {\square}(({\langle\rangle}{\exists}aNa\backslash Sb)/{\exists}aNa)]], {\blacksquare}Nt(s(n))\ \Rightarrow\ Sf
\using {\forall}L
\endprooftree
\justifies
[{\blacksquare}Nt(s(m))], [[{\square}(({\langle\rangle}{\exists}gNt(s(g))\backslash Sf)/{\exists}aNa), {\square}(({\langle\rangle}{\exists}gNt(s(g))\backslash Sf)/{\exists}aNa), \mbox{\fbox{${\forall}f{\forall}a((?{\blacksquare}(({\langle\rangle}Na\backslash Sf)/{\exists}bNb)\backslash {[]^{-1}}{[]^{-1}}(({\langle\rangle}Na\backslash Sf)/{\exists}bNb))/{\blacksquare}(({\langle\rangle}Na\backslash Sf)/{\exists}bNb))$}}, {\blacksquare}{\forall}a(({\langle\rangle}Na\backslash Sf)/({\langle\rangle}Na\backslash Sb)), {\square}(({\langle\rangle}{\exists}aNa\backslash Sb)/{\exists}aNa)]], {\blacksquare}Nt(s(n))\ \Rightarrow\ Sf
\using {\forall}L
\endprooftree
\justifies
[{\blacksquare}Nt(s(m))], [[{\square}(({\langle\rangle}{\exists}gNt(s(g))\backslash Sf)/{\exists}aNa), {\square}(({\langle\rangle}{\exists}gNt(s(g))\backslash Sf)/{\exists}aNa), \mbox{\fbox{${\blacksquare}{\forall}f{\forall}a((?{\blacksquare}(({\langle\rangle}Na\backslash Sf)/{\exists}bNb)\backslash {[]^{-1}}{[]^{-1}}(({\langle\rangle}Na\backslash Sf)/{\exists}bNb))/{\blacksquare}(({\langle\rangle}Na\backslash Sf)/{\exists}bNb))$}}, {\blacksquare}{\forall}a(({\langle\rangle}Na\backslash Sf)/({\langle\rangle}Na\backslash Sb)), {\square}(({\langle\rangle}{\exists}aNa\backslash Sb)/{\exists}aNa)]], {\blacksquare}Nt(s(n))\ \Rightarrow\ Sf
\using {\blacksquare}L
\endprooftree}
\end{center}}

\vspace{0.15in}

All this assigns the correct semantics:
\disp{
$[({\it Pres}\ ((\mbox{\v{}}{\it praise}\ {\it l})\ {\it j}))\wedge [({\it Pres}\ ((\mbox{\v{}}{\it like}\ {\it l})\ {\it j}))\wedge ({\it Fut}\ ((\mbox{\v{}}{\it love}\ {\it l})\ {\it j}))]]$}

\section{Cut elimination and decidability (proof idea)}
\label{Cutdec}
Cut elimination has several key steps and commutative steps. Here we consider only the key step
concerning the existential exponential modality. As usual, the proof proceeds by a double induction on
the size of the Cut formula and the sum of the heights  of the premises of the Cut occurrences. The so-called pseudo-key
step of a right application of \univexp{} or $\univexp_b$ (as left premise of Cut) and a contraction (as right premise of Cut)
is more involved but still standard.\footnote{Recall that both 
\univexp{} and $\univexp_b$ allow only contraction. No weakening nor expansion are associated to these
connectives.} Notice that crucially, the $\univexp_b$-contraction must be defined for $\univexp_b$-modalized sequences as is the case
in Figure~\ref{slunivexpb}.\footnote{The so-called full Lambek calculus with contraction enjoys Cut-elimination if the contraction
rule is generalized to sequences of types.}

The  key Cut steps involve the structural rules $\univexp C$ and $\exstexp M$. The case of $\univexp C$ is standard in the literature 
of linear logic; we therefore omit it. What is really new is the $\univexp M$ key Cut, which is as follows.
(This key step simply does not exist in the case of the calculi
$\DbCb$, nor in $\DbCbMr$ because there are only succedent occurrences of the existential exponential.)
Where $\univexp\Delta(\Gamma_i)=\univexp\Delta_1,\Gamma_i,\univexp\Delta_2$, we have
that the following rule $\exstexp GM$:\\
\disp{$
\prooftree
\univexp\Delta(\Gamma_1)\yields\exstexp A
\tb
\univexp\Delta(\Gamma_2)\yields\exstexp A
\justifies
\univexp\Delta(\Gamma_1,\Gamma_2)\yields\exstexp A
\using\exstexp GM
\endprooftree
$\label{GMrule}}
is derivable from $\exstexp M$ by application of $\exstexp M$ and the permutation
and contraction $\univexp$-steps without the use of Cut. Then there is the key step:\\

$
\prooftree
\prooftree
\univexp\Delta(\Gamma_1)\yields \exstexp A
\tb
\univexp\Delta(\Gamma_2)\yields \exstexp A
\justifies
\univexp\Delta(\Gamma_1,\Gamma_2)\yields \exstexp A
\using 
\exstexp GM
\endprooftree
\tb
\prooftree
\justifies
\univexp\Theta(\exstexp A)\yields \exstexp B
\using \exstexp L
\endprooftree
\justifies
\univexp\Theta(\univexp\Delta(\Gamma_1,\Gamma_2))\yields \exstexp B
\using Cut
\endprooftree
$\\\ \\

$\leadsto$
$
\prooftree
\prooftree
\univexp\Delta(\Gamma_1)\yields\exstexp A
\tb
\univexp\Theta(\exstexp A)\yields \exstexp B
\justifies
\univexp\Theta(\univexp\Delta(\Gamma_1))\yields \exstexp B
\using Cut
\endprooftree
\tb
\prooftree
\univexp\Delta(\Gamma_2)\yields\exstexp A
\tb
\univexp\Theta(\exstexp A)\yields \exstexp B
\justifies
\univexp\Theta(\univexp\Delta(\Gamma_2))\yields \exstexp B
\using Cut
\endprooftree
\justifies
\univexp\Theta(\univexp\Delta(\Gamma_1,\Gamma_2))\yields \exstexp B
\using ?GM
\endprooftree
$\\\ \\

\noindent Let us see now the proof that the generalized Mingle rule for \exstexp{} is Cut-free derivable in \DbCM{} using
\univexp{}-contractions and $\exstexp$-Mingle. If we write $\univexp\Delta(\Sigma)$ as $\univexp\Delta_1,\Sigma,\univexp\Delta_2$
for arbitrary configurations $\Delta_i$ and $\Sigma$, we have the following $\exstexp$-Mingle derivation:
$$
\prooftree
\univexp\Delta(\Gamma_1)\yields\exstexp A
\tb
\univexp\Delta(\Gamma_1)\yields\exstexp A
\justifies
\mathcal{S}:=\univexp\Delta(\Gamma_1),\univexp\Delta(\Gamma_2)\yields\exstexp A
\using\exstexp M
\endprooftree
$$
To the end-sequent $\mathcal{S}$ of the above derivation we apply a finite number of \univexp{}-permutation steps and we get the provable sequent:
$$\univexp\Delta_1,\univexp\Delta_1,\Gamma_1,\Gamma_2,\univexp\Delta_2,\univexp\Delta_2\yields\exstexp A$$
Finally, to the above sequent we apply a finite number of \univexp{}-contraction steps obtaining:
$$\univexp\Delta_1,\Gamma_1,\Gamma_2,\univexp\Delta_2\yields\exstexp A$$ 
This last sequent can be written as:
$$\univexp\Delta(\Gamma_1,\Gamma_2)\yields\exstexp A$$
\noindent
Hence, the $\exstexp$-GM structural rule is Cut-free derivable in \DbCM{}.
This ends the proof idea
of the Cut admissibility of the four calculi we have considered.

In order to prove the decidability of a linguistically sufficient special case of \DbCbMr{}, which we call \emph{polar bracket non-negative\/} \DbCbMr{}
we introduce two useful technical tools: \emph{bracket-count} of a type and
\emph{degree of contraction} of a sequent $\mathcal{S}$.
Building upon (\cite{vsm:add}), we define the bracket-count of a sequent recursively as follows:
\defn{Bracket-count}{
Where $A$ and $B$ are arbitrary \DbCbMr{}-types:
$$
\begin{array}{lll}
\#_{[]}(A)&=&0\mbox{ if $A$ is atomic}\\
\#_{[]}(A\product B)&=&\#_{[]}(A)+\#_{[]}(B)\\
\#_{[]}(A\swprod{i} B)&=&\#_{[]}(A)+\#_{[]}(B)\\
\#_{[]}(B/A)&=&\#_{[]}(B)-\#_{[]}(A)\\
\#_{[]}(B\scircum{k} A)&=&\#_{[]}(B)-\#_{[]}(A)\\
\#_{[]}(A\bsl B)&=&\#_{[]}(B)-\#_{[]}(A)\\
\#_{[]}(B\sinfix{k} A)&=&\#_{[]}(B)-\#_{[]}(A)\\
\#_{[]}(\mybrack A)&=&\#_{[]}(A)+1\\
\#_{[]}(\abrack A)&=&\#_{[]}(A)-1\\
\#_{[]}(\univexp A)&=&\#_{[]}(A)\\
\#_{[]}(\exstexp A)&=&\#_{[]}(A)
\end{array}
$$
Where $\Delta,\Delta_i\mbox{ ($i=1,	\cdots, n,$ $n>0$)}$ are \DbCbMr{}-configurations:
$$
\begin{array}{lll}
\#_{[]}(\Lambda)&=&0\\
\#_{[]}(A,\Delta)&=&\#_{[]}(A)+\#_{[]}(\Delta)\\
\#_{[]}(\sep)&=&0\\
\#_{[]}(A\{\Delta_1:\cdots:\Delta_n\},\Delta)&=&\displaystyle \sum_{i=1}^n \#_{[]}(\Delta_i)+\#_{[]}(\Delta)\\
\#_{[]}([\Delta])&=&\#_{[]}(\Delta)+1\
\end{array}
$$
\label{defbracketcount}}
\defn{Degree of Contraction}{
We define the \emph{degree of contraction} of a sequent $\mathcal{S}$$:=$$\Delta\yields A$, $\degcon(\mathcal{S})$, in terms of bracket counts as follows:
$$\degcon(\mathcal{S})\bydef\#_{[]}(A)-\#_{[]}(\Delta)$$
\label{defdegcon}
}
We see now some simple facts on the degree of contraction of sequents:\\

\noindent -\textbf{ Fact 1}: Given a derivation whose last rule is a binary or unary bracket rule with conclusion
$\mathcal{S}$ and premises $\mathcal{S}_i$:
$$\degcon(S)\geq\degcon(S_i)$$
\noindent -\textbf{ Fact 2}: Suppose that the last rule of a derivation is the contraction rule where the configuration $\univexp\Gamma$ is a bracket-free configuration:
$$
\prooftree
\mathcal{S}_2:=\Delta\langle\univexp\Gamma,[\univexp\Gamma,\Theta]\rangle\yields A
\justifies
\mathcal{S}_1:=\Delta\langle\univexp\Gamma,\Theta\rangle\yields A
\using\univexp C_b
\endprooftree
$$
\noindent Then we have:
$$\degcon(\mathcal{S}_1) > \degcon(\mathcal{S}_2)$$
\noindent -\textbf{ Fact 3}: Suppose that the last rule of a derivation is the restricted Mingle rule, where all type-occurrences are bracket-free:
$$
\prooftree
\mathcal{S}_2:=\Delta_1\yields A\tb
\mathcal{S}_3:=\Delta_2\yields \exstexp A
\justifies
\mathcal{S}_1:=\Delta_1,\Delta_2\yields \exstexp A
\using\exstexp M_r
\endprooftree
$$
\noindent Then we have: 
$$\degcon(\mathcal{S}_1) \geq \degcon(\mathcal{S}_2)+\degcon(\mathcal{S}_3)$$

\noindent Finally, a useful arithmetic tool is the length of an arbitrary sequent $\mathcal{S}:=\Delta\yields A$, $|\mathcal{S}|$. The
well known length of a type, which is simply its number of connectives, and the (overloaded) length of a configuration
$\Delta$, $|\Delta|$, which is the sum of the lengths of all its type-occurrences, we define $|\mathcal{S}|$ as $|\Delta|+|A|$.
We have the following theorem:
\disp{
The Cut-free proof-search space in \DbCbMr{} is finite.
\label{thmproofsearchfinite}}
\prf{
Let \lexord{} be the total strict lexicographical order in $\mathbb{N}^2$. Consider a sequent $\mathcal{S}$ such that $\degcon(\mathcal{S})\geq 0$ (for otherwise it could not be provable).
We want to check its provability. We can expand the current goal sequent $\mathcal{S}$ of the proof-search 
space $\mathbf{ProofSearch}$ by a finite number of goal sequents, 
which can be either the subgoals of a logical rule or a structural rule. We associate to each sequent $\mathcal{S}$ of $\mathbf{ProofSearch}$ its 
\emph{measure} $\mu(\mathcal{S})\bydef(\degcon(\mathcal{S}),|\mathcal{S}|)$. If we expand $\mathcal{S}$ with a contraction rule, the degree of
contraction is strictly decreased. In case of a restricted Mingle rule or a logical rule the degree of contraction may be decreased or remain equal.
In case that the degree of contraction remains equal, the lengths of the premises of the applied rule are strictly decreased. Hence, $\mathbf{ProofSearch}$
is a finitely branched tree such that any path $(S_i)_{i>0}$ of it satisfies $\mu(\mathcal{S}_{i+1}) \lexord \mu(\mathcal{S}_{i})$ for all $i$. Since \lexord{}
is well-founded every strictly decreasing sequence is finite. Therefore, by K\"{o}nig's lemma, $\mathbf{ProofSearch}$ is finite.}

From the preceding theorem, it follows that \DbCbMr{} is decidable in the case that the exponential subtypes are bracket-free in
the sense of not containing bracket modalities within exponentials which give rise to antecedent antibracket modalities nor succedent bracket modalities . We call the restriction to such types polar bracket non-negative
\DbCbMr{}.

Whether the calculus \DbCM{} is decidable is an open problem. However, it is interesting to notice that \DbCM{} extended with additive connectives
is undecidable. In fact, the Lambek Calculus with additives and the connective \univexp{}, of which \DbCM{} with additives is a conservative
extension, is already undecidable. This can be proved by a Girard-style translation $(\cdot)^\bullet$ between the full Lambek calculus with contraction
(\FLC{}) and the full Lambek calculus with \univexp{}-contraction (\FLexpC{}) as follows:
$$
\begin{array}{lll}
A^\bullet&=&A\mbox{ if $A$ is atomic}\\
(B/A)^\bullet&=&B^\bullet/\univexp A^\bullet\\
(A\bsl B)^\bullet&=&\univexp A^\bullet\bsl B^\bullet\\
(A\oplus B)^\bullet&=&\univexp A^\bullet\oplus\univexp B^\bullet\\
(A\& B)^\bullet&=&A^\bullet\&B^\bullet\\
(\Delta\yields A)^\bullet&=&\univexp\Delta^\bullet\yields A^\bullet
\end{array}{}
$$

We can prove the following theorem:
\thm{Embedding translation between \FLC{} and \mbox{\bf \FLC\rm !}}
{$\FLC\vdash\Delta\yields A\mbox{ iff }\FLexpC(\Delta\yields A)^\bullet$
\label{embeddingthm}
}
\corol{Undecidability of \mbox{\bf \FLC\rm !}}{
It has been proved that \FLC{} is undecidable \cite{kareletal:flcundec}. If \FLexpC{} were decidable, for any \FLC{}-sequent  $\Delta\yields A$,
we could decide whether its translation $(\Delta\yields A)^\bullet$ is provable. We would have then that \FLC{} is decidable. Contradiction.
}
\section*{Appendix: $\exstexp$-Mingle vs.\ $\exstexp$-Expansion}
Consider the following structural rule called $?$-expansion. For any type $A$:
\disp{
\prooftree
\Delta\langle?A\rangle\yields B
\justifies
\Delta\langle?A,?A\rangle\yields B
\using E
\endprooftree
}
It is straightforward to see that $\DbCM+Cut$ is deductively equivalent to $\DbCM+Exp-Mingle+Cut$. However,  $\DbCM+Cut$ enjoys Cut elimination,
but $\DbCM+Exp-Mingle+Cut$ does not enjoy Cut elimination.
%
%

\bibliographystyle{plain}
\bibliography{bib150311}

\end{document}